\documentclass{emulateapj}

\slugcomment{Accepted for publication in ApJL}

\begin{document}

\title{The high albedo of the hot Jupiter Kepler-7\,b}

\author{Brice-Olivier Demory\altaffilmark{1}, Sara Seager\altaffilmark{1}, Nikku Madhusudhan\altaffilmark{2}, Hans Kjeldsen\altaffilmark{3}, J{\o}rgen~Christensen-Dalsgaard\altaffilmark{3}, Micha{\"e}l Gillon\altaffilmark{4}, Jason F. Rowe\altaffilmark{5}, William F. Welsh\altaffilmark{6}, Elisabeth R. Adams\altaffilmark{7}, Andrea Dupree\altaffilmark{7}, Don McCarthy\altaffilmark{8}, Craig Kulesa\altaffilmark{8}, William J. Borucki\altaffilmark{5}, David~G. Koch\altaffilmark{5} and the \textit{Kepler} Science Team}
\altaffiltext{1}{Department of Earth, Atmospheric and Planetary Sciences, Massachusetts Institute of Technology, 77 Massachusetts Ave., Cambridge, MA 02139, USA. demory@mit.edu}
\altaffiltext{2}{Department of Astrophysical Sciences, Princeton University, Princeton, New Jersey 08544, USA.}
\altaffiltext{3}{Department of Physics and Astronomy, Aarhus University, DK-8000 Aarhus C, Denmark.}
\altaffiltext{4}{Institut d'Astrophysique et de G\'eophysique, Universit\'e de Li\`ege, All\'ee du 6 Ao\^ut, 17, Bat. B5C, Li\`ege 1, Belgium.}
\altaffiltext{5}{NASA Ames Research Center, Moffett Field, CA 94035, USA.}
\altaffiltext{6}{Astronomy Department, San Diego State University, San Diego, CA 92182, USA.}
\altaffiltext{7}{Smithsonian Astrophysical Observatory, 60 Garden St., Cambridge, MA 02138, USA}
\altaffiltext{8}{Steward Observatory, University of Arizona, 933 N. Cherry Ave, Tucson, AZ 85721, USA}

\begin{abstract}
Hot Jupiters are expected to be dark from both observations (albedo
upper limits) and theory (alkali metals and/or TiO and VO
absorption). However, only a handful of hot Jupiters have been
observed with high enough photometric precision at visible
wavelengths to investigate these expectations. The NASA \textit{Kepler} mission provides a means to
widen the sample and to assess the extent to which hot Jupiter
albedos are low.  We present a global analysis of Kepler-7\,b based on
Q0-Q4 data, published radial velocities, and asteroseismology
constraints. We measure an occultation depth in the \textit{Kepler}
bandpass of 44$\pm$5 ppm. If directly related to the albedo, this translates to a
\textit{Kepler} geometric albedo of 0.32$\pm$0.03, the most precise
value measured so far for an exoplanet. We also characterize the planetary
orbital phase lightcurve with an amplitude of $42\pm$4 ppm. Using atmospheric models, 
we find it unlikely that the high albedo is due to a dominant thermal component and
propose two solutions to explain the observed planetary flux.
Firstly, we interpret the Kepler-7\,b albedo as resulting from an excess reflection over 
what can be explained solely by Rayleigh scattering, along with a nominal 
thermal component. This excess reflection might indicate the presence of a 
cloud or haze layer in the atmosphere, motivating new modeling and observational efforts. 
Alternatively, the albedo can be explained by Rayleigh
scattering alone if Na and K are depleted in the atmosphere by a
factor of 10-100 below solar abundances. 

\end{abstract}


\section{Introduction}

More than 30 hot Jupiters benefit from observations of their
emitted radiation from near to mid-infrared, where the measurement of
their thermal emission is the most favorable. \textit{Spitzer} made a
significant contribution by producing a flurry of results allowing us
to derive general properties of hot-Jupiter atmospheres
\citep{DemingSeager2009}. Those planets are strongly irradiated by
their host stars and their equilibrium temperatures were early
estimated to be above 1000K \citep{SeagerSasselov2000}.
Those observations confirm that hot Jupiters efficiently reprocess the 
incident stellar flux into thermal reemission, exhibiting low flux at visible 
wavelengths \citep{Marley1999, SeagerWhitney2000, Sudarsky2003}.

Characterization of transiting hot-Jupiter reflected light suffers from the scarcity of 
observations. The planetary to stellar flux ratio is of the order of $10^{-5}$ 
in the visible, two orders of magnitude less than mid-infrared signatures. 

To date, eleven planets have an upper limit constraint on their geometric albedo :
$\tau$Boo\,b \citep{Charbonneau1999,Leigh2003a,Rodler2010}, $\upsilon$And\,b \citep{Cameron2002}, 
HD75289A\,b \citep{Leigh2003b,Rodler2008}, HD209458b \citep{Rowe2008}, CoRoT-1b \citep{Alonso2009,Snellen2009},
CoRoT-2b \citep{Alonso2010,Snellen2010}, HAT-P-7b \citep{Christiansen2010,Welsh2010},
Kepler-5b \citep{Kipping2010,Desert2011}, Kepler-6b \citep{Kipping2010,Desert2011},
Kepler-7b \citep{Kipping2010} and HD189733b \citep{Berdyugina2011}.

Eight of them corroborate early theoretical predictions: with $A_g<0.3$ (3$\sigma$ upper limit) 
hot Jupiters are dark in the visible. However \citet{Cameron2002} determined $A_g < 0.42$ (3$\sigma$) from spectroscopy 
in the 380-650nm range for $\upsilon$And\,b, \citet{Kipping2010} reported $A_g = 0.38\pm0.12$\footnote{Determined in the \textit{Kepler} bandpass, which has a $>$5\% response between 423 and 897 nm \citep{Koch2010}} for Kepler-7\,b and \citet{Berdyugina2011} 
determined a $V$-band albedo of $A_g=0.28\pm0.16$ for HD189733b from polarimetry, suggesting dominance of reflected light over thermal emission.

Solar system giant planets have geometric albedos of 0.32 (Uranus) to 0.50 (Jupiter) in a bandpass 
similar to \textit{Kepler}'s \citep{Karkoschka1994}. 
Those objects harbor bright cloud decks made of ammonia and 
water ice that are highly reflective at visible wavelengths. In contrast to the solar system 
giant planets, atmosphere models show that the presence of alkali metals in hot-Jupiter 
atmospheres (Na and K) as well as TiO and VO (at the hotter range) causes significant 
absorption at visible wavelengths.

We report in this letter the characterization of the hot Jupiter Kepler-7\,b 
\citep{Latham2010}, based on \textit{Kepler} Q0-Q4 data. We present the photometry and 
data analysis in Sect.~2, while corresponding results are shown in Sect.~3.
Discussion and atmospheric analysis are finally presented in Sect.~4. 

\section{Observations and Data Analysis}

\subsection{Kepler photometry}

Kepler-7\,b belongs to the first set of new planets published by the \textit{Kepler} science team in early 2010. Kepler-7\,b is a 4.8-day period hot Jupiter orbiting a $V$=13.04 sub-giant G star with $M_\star=1.347\pm0.07M_{\odot}$ and $R_\star=1.843\pm0.07R_{\odot}$ \citep{Latham2010}. Like all objects located in the \textit{Kepler} field, Kepler-7 benefits from nearly continuous photometric monitoring since mid-2009.

We base our analysis on the Q0-Q4 quarters, which represent nearly one year of observations. Data recorded during each quarter are differentiated in short- and long-cadence timeseries, that are binnings per 58.84876s and 29.4244min respectively of the same CCD readouts. Five long-cadence \citep{Jenkins2010} and six short cadence \citep{Gilliland2010} datasets are used as part of this study, representing 272,719 photometric datapoints and 311.68 effective days of observations, out of which 175.37 days have also been recorded in short cadence. We used the raw photometry for our purpose.

Kepler-7 is a photometrically quiet star: apart from the 4.88-day period transit signals, no evidence of significant stellar variability is apparent in the data.

\subsection{Data analysis}

For the purpose of this global analysis, we used the implementation of the Markov Chain Monte-Carlo (MCMC) algorithm presented in \citet{Gillon2009,Gillon2010}. MCMC is a Bayesian inference method based on stochastic simulations that samples the posterior probability distributions of adjusted parameters for a given model. Our MCMC implementation uses the Metropolis-Hastings algorithm to perform this sampling. Our nominal model is based on a star and a transiting planet on a Keplerian orbit about their center of mass.

Our global analysis was performed using 199 transit and occultation lightcurves in total, out of which 70 were acquired in short cadence mode. We discarded 13 lightcurves because of discontinuities due to spacecraft roll, change of focus, pointing offsets or safe mode events. Input data also include the 9 radial velocity points obtained from NOT/FIES (FIber-fed Echelle Spectrograph) that were published in \citet{Latham2010}.

As the focus of this study is on using the transits and occultations to refine the system parameters, for the model fitting we use only the photometry near the eclipse events. Windows of width 0.6 days (12.3\% of the orbit) surrounding eclipses were used to measure the local out-of-transit baseline, while minimizing the computation time. A dilution of 2.7$\pm$0.5\% was determined from MMT/ARIES\footnote{MMT is a joint facility of the Smithsonian Institution and University of Arizona.} (ARizona Infrared imager and Echelle Spectrograph) observations and applied to both transit and occultation photometry.

The excellent sampling of the transit lightcurve motivated us to fit for the limb-darkening (LD) coefficients. For this purpose, we assumed a quadratic law and used $c_1=2u_1+u_2$ and $c_2=u_1-2u_2$ as jump parameters, where $u_1$ and $u_2$ are the quadratic coefficients. 

The MCMC has the following set of jump parameters: the planet/star flux ratio, the impact parameter $b$, the transit duration from first to fourth contact, the time of minimum light, the orbital period, $K'=K\sqrt{1-e^2}P^{1/3}$, where $K$ is the radial-velocity semi-amplitude, the occultation depth, the two LD combinations $c_1$ and $c_2$ and the two parameters $\sqrt{e}\cos\omega$ and $\sqrt{e}\sin\omega$. A uniform prior distribution is assumed for all jump parameters.

\subsubsection{Model and systematics}
The transit and occultation photometry are modeled with the \citet{MandelAgol2002} model, multiplied by a second order polynomial accounting for stellar and instrumental variability. 
Baseline model coefficients are determined for each lightcurve with the SVD method \citep{Press1992} at each step of the MCMC. 
Correlated noise was accounted for following \citet{Winn2008,Gillon2010}, to ensure reliable error bars on the fitted parameters. For this purpose, we compute a scaling factor based on the standard deviation of the binned residuals for each lightcurve with different time bins. The error bars are then multiplied by this scaling factor. We obtained a mean scaling factor of 1.04 for all photometry, denoting a negligible contribution from correlated noise. The mean global photometric RMS per 30-min bin is 96 parts per million (ppm).

\subsubsection{Asteroseismology}
The data series for Kepler-7 contains 9 months of data at a cadence
of 1 minute. The power spectrum shows a clear excess of power near 1.05mHz. 
The asteroseismic analysis of the data was performed using the
pipeline developed at the Kepler Asteroseismic Science Operations Center as
described in detail by \citet{Christensen2008,Christensen2010,Huber2009,
Gilliland2011}. Using the matched filter approach we determine a value
for the large separation of 56$\mu$Hz. Locating the asymptotic frequency
structure in the folded power allows a robust identification of 13 individual p-mode 
frequencies and an estimate of the scatter on those frequencies (0.9$\mu$Hz).
The frequencies resulting from this analysis were fitted to stellar models in the same
manner as in  \citet{Christensen2010}, using also the effective
temperature and metallicity ([Fe/H]) determined by \citet{Latham2010}.
The models did not include diffusion and settling.
Models were computed without overshoot from the convective core, as well as with 
overshoot of $0.1H_p$ and $0.2H_p$, where $H_p$ is the pressure scale height at the
edge of the convective core.
The observed frequencies and effective temperature were fitted to the models
in a least-squares sense, resulting in a weighted average of the stellar properties.
Interestingly, only models with overshoot provided acceptable fits to the frequencies 
within the observed range of the effective temperature.

We used the resulting stellar density (see Table 1) as a Bayesian prior in the MCMC and the corresponding stellar mass to derive the system's physical parameters.

\subsubsection{Phase curve}
About 312 days of Kepler-7 observations are covered in Q0-Q4 data. This motivated us to search for the planetary phase signature. We first removed the transits and fitted the long and short cadence data to remove temporal linear trends. Stellar and instrumental induced modulation on the photometry was then removed by pre-whitening the raw data using \texttt{Period04} software \citep{Lenz2005}. This step allowed us to filter out all frequencies below the orbital frequency and those that are not connected with the planetary orbit period.

\section{Results}

\begin{deluxetable*}{ll}
\tabletypesize{\scriptsize}
\tablecaption{Kepler-7 system parameters}
\tablenum{1}
\tablehead{\colhead{Parameters} & \colhead{Value} }
\startdata
\textit{Jump parameters} &   \\
\tableline
 &   \\
Planet/star area ratio $(R_p/R_s)^2$ & $0.006772^{+0.000018}_{-0.000021}$ \\
$b'=a \cos i /R_{\star}$ [$R_{\star}$] & $0.5565^{+0.0060}_{-0.0063}$ \\
Transit width [d] & $0.21777^{+0.00023}_{-0.00021}$ \\
$T_0$ - 2450000 [HJD] & $4967.27599^{+0.00019}_{-0.00020}$ \\
Orbital period $P$ [d] & $4.8854830^{+0.0000042}_{-0.0000041}$  \\
RV $K'$ [m\,s$^{-1}$\,d$^{1/3}$]  & $73.1^{+6.7}_{-6.8}$  \\
$\sqrt{e} \cos \omega$ & $0.0376^{+0.0143}_{-0.0153}$  \\
$\sqrt{e} \sin \omega$ & $-0.0016^{+0.0026}_{-0.0022}$  \\
$c_1 = 2 u_1 + u_2$ & $0.922^{+0.011}_{-0.011}$  \\
$c_2 = u_1 - 2 u_2$ & $-0.143^{+0.022}_{-0.022}$  \\
Occultation depth & $0.000044^{+0.000005}_{-0.000006}$  \\
 &    \\
\textit{Deduced stellar parameters} &    \\
\tableline
 &    \\
$u_1$ & $0.344^{+0.007}_{-0.007}$  \\
$u_2$ & $0.232^{+0.009}_{-0.009}$  \\
Density $\rho_{star}$ [g\, cm$^{-3}$] & $0.231^{+0.004}_{-0.004}$  \\
Surface gravity $\log g_*$ [cgs] & $3.960^{+0.005}_{-0.005}$  \\
Mass $M_{\star}$ [$M_{\odot}$]& $1.36^{+0.03}_{-0.03}$  \\
Radius $R_{\star}$ [$R_{\odot}$]& $2.02^{+0.02}_{-0.02}$  \\
 &    \\
\textit{Asteroseismic parameters} &    \\
 \tableline
  &    \\
Density $\rho_{star}$ [g\, cm$^{-3}$] &  0.252 $\pm$   0.003    \\
Mass $M_{star}$ [$M_{\odot}$]&       1.359  $\pm$  0.031     \\
Radius $R_{star}$ [$R_{\odot}$] &       1.966  $\pm$ 0.013    \\
Stellar age [Gyr] &       3.3  $\pm$  0.4    \\
Surface gravity $\log g_*$ [cgs]  &       3.984  $\pm$  0.006    \\
&  \\
\textit{Deduced planet parameters}  &  \\
\tableline
 &  \\
RV $K$  [m\,s$^{-1}$]  & $43.1^{+3.9}_{-4.0}$ \\
$b_{transit}$ [$R_{\star}$]  & $0.557^{+0.006}_{-0.006}$ \\
$b_{occultation}$ [$R_{\star}$]  & $0.556^{+0.006}_{-0.006}$ \\
$T_{occultation}$ - 2450000 [HJD]  & $4974.6086^{+0.0040}_{-0.0028}$ \\
Orbital semi-major axis $a$ [AU]  & $0.06246^{+0.00046}_{-0.00046}$ \\
Orbital inclination $i$ [deg]  & $85.18^{+0.076}_{-0.074}$ \\
Orbital eccentricity $e$  & $0.001^{+0.001}_{-0.001}$ \\
Argument of periastron $\omega$ [deg]  & $357.1^{+4.4}_{-2.8}$ \\
Density $\rho_{P}$ [g\, cm$^{-3}$]  &$0.14^{+0.01}_{-0.01}$ \\
Surface gravity log $g_{P}$ [cgs]  & $2.62^{+0.04}_{-0.04}$ \\
Mass $M_{P}$ [$M_{Jup}$]  & $0.443^{+0.041}_{-0.042}$ \\
Radius $R_{P}$ [$R_{Jup}$]  & $1.614^{+0.015}_{-0.015}$ \\
\enddata
\end{deluxetable*}

We present Kepler-7\,b's system parameters in Table 1. Each value is the median of the marginal posterior distribution obtained for the relevant parameter. Error bars are the corresponding 68.3\% probability interval. Figure 1 shows the phase-folded transit photometry of Kepler-7\,b.

\begin{figure*}
\begin{center}
\epsscale{1.0}
\plotone{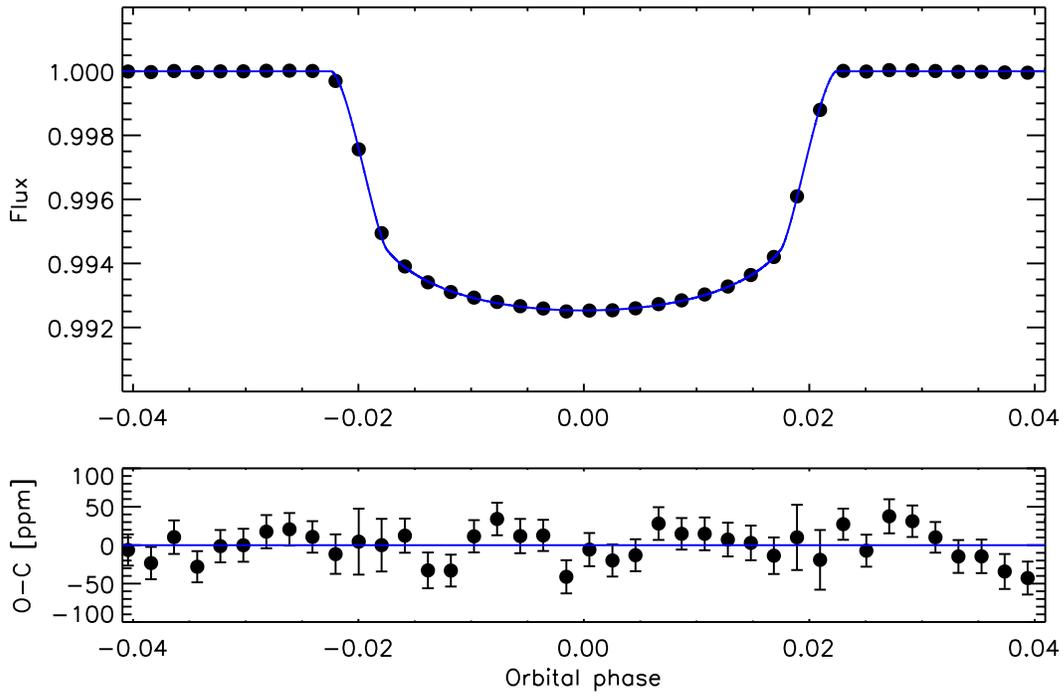}
\caption{Top : Kepler-7\,b phase-folded transit lightcurve with best-fit model superimposed. Binned per 15min. Error bars are smaller than the plotted datapoints. Bottom : residuals.\label{fig1}}
\end{center}
\end{figure*}

Our results confirm the very low density of Kepler-7\,b. The external constraint from asteroseismology modeling causes the planet radius to significantly increase as compared to the discovery paper \citep{Latham2010}. Our MCMC analysis yields a planetary radius of $R_P=1.61\pm0.02R_{Jup}$ and a mass of $M_P=0.44\pm0.04M_{Jup}$, which gives a surprisingly low density of $\rho_P=0.14\pm0.01$ g\, cm$^{-3}$. Interestingly, Kepler-7b's properties are close to WASP-17b's \citep{Anderson2009}, the least dense transiting planet discovered so far with $\rho_P=0.12\pm0.06$ g\, cm$^{-3}$. Both planets are of similar mass and orbit evolved stars.

We find a marginal orbital eccentricity of $e=0.001\pm0.001$.

We determine an occultation depth of $44\pm$5 ppm. The corresponding phase-folded lightcurve is shown on bottom panel of Fig.~2. 

Finally we find an orbital phase curve of $42\pm$4 ppm amplitude that is consistent with the occultation depth and phased with Kepler-7\,b's transits and occultations. We model the phase curve assuming a Lambert law phase-dependent flux-ratio \citep{Sobolev1975} :
\begin{displaymath} 
\phi(\alpha)=A_{g}\left(\frac{R_{p}}{a}\right)^{2}\left[\frac{\sin(\alpha)+(\pi-\alpha)\cos(\alpha)}{\pi}\right]
\end{displaymath}
where $\alpha$ is the orbital phase, $A_{g}$ is the geometric albedo, $R_{p}$ is the planetary radius and $a$ is the orbital semi-major axis. A value of $A_{g}=0.31\pm0.03$ is deduced from the phase curve using $R_{p}$ and $a$ values from Table 1. Although a perfectly reflecting Lambertian sphere has $A_{g}=\frac{2}{3}$,  the Lambert law can represent a scatterer which takes on a lower $A_{g}$ value for an atmosphere with some absorption \citep{Seager2010}. The resulting lightcurve is shown on top panel of Fig.~2, with the best-fit model superimposed. No ellipsoidal variations are detected ($-1\pm3$ppm).

\begin{figure*}
\begin{center}
\epsscale{1.0}
\plotone{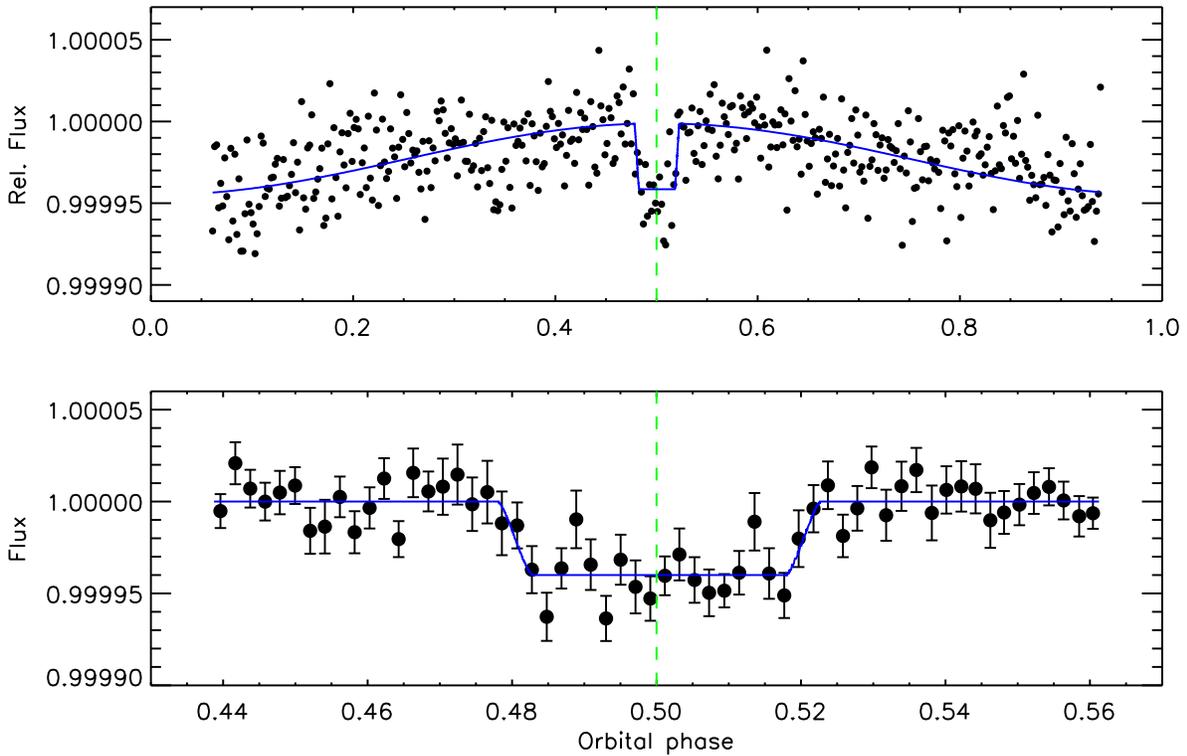}
\caption{Top : Kepler-7\,b orbital phase curve with best-fit model (see text) superimposed. Transits are omitted. Bottom : Kepler-7\,b phase-folded occultation lightcurve with best-fit model. Binned per 15min.\label{fig2}}
\end{center}
\end{figure*}

\section{The High Albedo of Kepler-7\,b}
If directly related to the albedo, the measured occultation depth of 44$\pm$5 ppm translates to a 0.32$\pm$0.03 geometric albedo as measured by \textit{Kepler}. The \textit{Kepler} bandpass encompasses a large range of wavelengths, from 0.4 to 0.9 $\mu$m. Albedo values reported in this paper for Kepler-7\,b are averaged over this spectral domain. For the highly irradiated exoplanets, a significant part of the thermal emission leaks into the red end of this bandpass, making the occultation depth larger. With a 4.9-d long orbit, Kepler-7\,b would not be expected to be one of the hottest giant planets found to date. However, its host star with a $T_{\rm eff}$=5933K, 1.4 $M_\odot$ and 2.0 $R_\odot$, is about 4.5 times more luminous than the Sun. These compensating factors make it necessary to estimate the relative contributions of thermal emission and reflected light to the occultation depth.

The possible relative contributions of thermal emission and reflected light to the observed flux in the {\it Kepler} bandpass are shown in Fig.~3. The thermal emission is represented by an effective brightness temperature ($T_B$), and the reflected light by the geometric albedo ($A_g$). Also shown are the dayside equilibrium temperatures corresponding to atmosphere with efficient versus inefficient energy redistribution (dotted lines). The degeneracy between $T_B$ and $A_g$ is evident. The observed occultation depth allows for geometric albedos as high as 0.35 for $T_B =$ 1500K in the {\it Kepler} bandpass. On the other hand, allowing for a zero geometric albedo requires an extremely high $T_B$ of 2500-2600K, which is $\sim$ 400K higher than the maximum equilibrium temperature. Completely breaking the degeneracy between $A_g$ and $T_B$ requires additional observations in the visible and near-infrared. However, tentative constraints can be placed on the various sources of opacity and scattering using a physical model atmosphere. 

\begin{figure}
\begin{center}
\epsscale{1.2}
\plotone{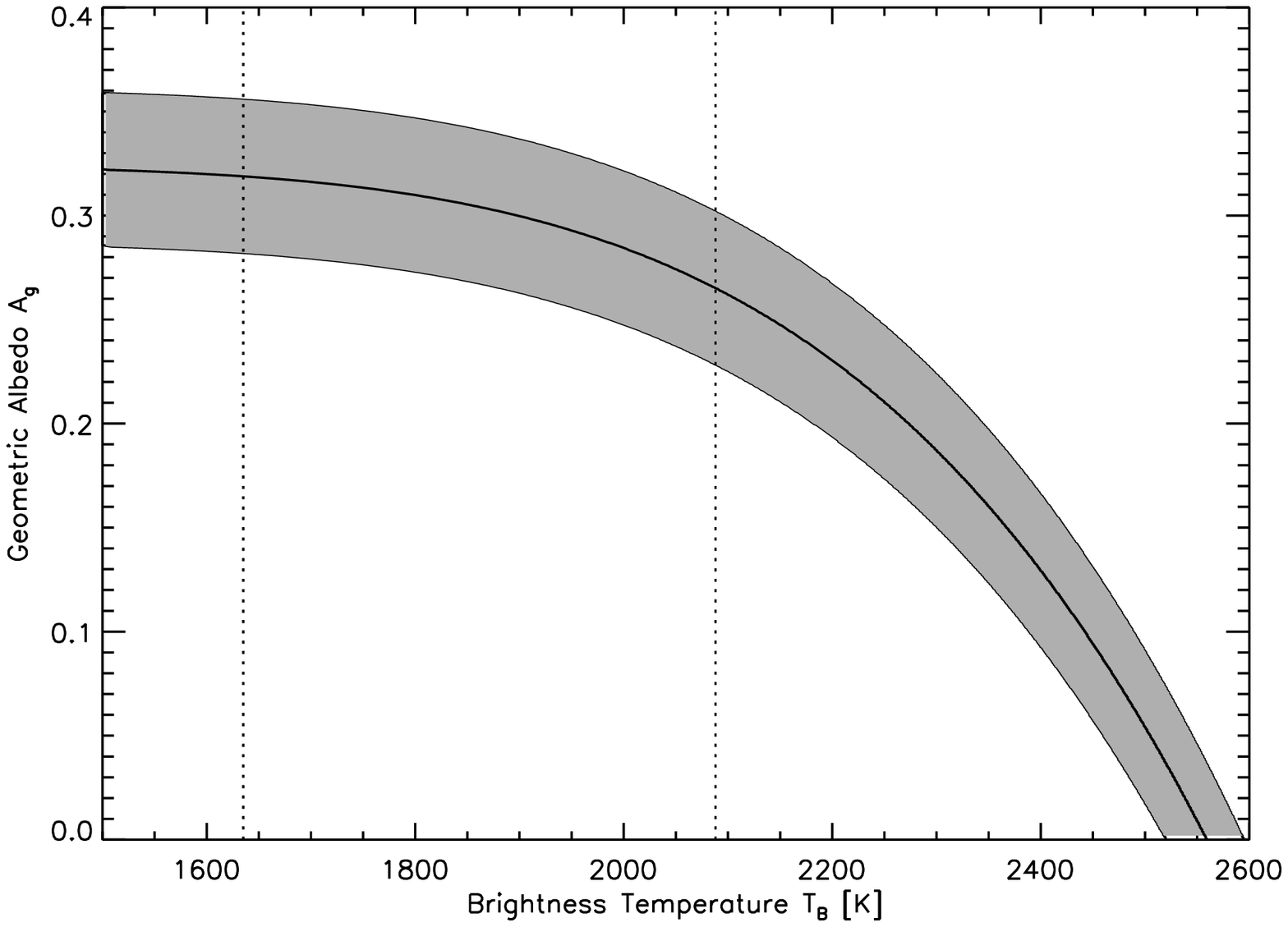}
\caption{Model-independent domain of allowed geometric albedo $A_g$ (reflected light) and brightness temperature $T_B$ (thermal emission) combinations, constrained by the observed \textit{Kepler} occultation depth. The shaded area represents the 1$\sigma$ confidence domain found using a Kurucz spectrum for the star and blackbody emission for the planet. The dotted lines depict the equilibrium temperature assuming full redistribution (left) and no redistribution (right).\label{fig3}}
\end{center}
\end{figure}

The thermal and reflection spectra of the planetary atmosphere depend on the various atomic and molecular opacities, sources of scattering, and the temperature structure, all of which constitute a large number of free parameters under-constrained by the single data point available. Nevertheless, the precise measurement allows us to constrain regions of the parameter space, given plausible assumptions about energy balance and atmospheric chemistry. We use the exoplanet atmospheric modeling and retrieval technique of \citet{Madhusudhan2009}. The model computes line-by-line radiative transfer in a plane-parallel atmosphere, with the assumption of hydrostatic equilibrium and global energy balance, and parametrized chemical composition and temperature structure. A key aspect of the model is the flexibility to explore a wide range of molecular abundances and pressure-temperature ($P$-$T$) profiles, without any assumptions of chemical or radiative equilibrium. We consider all the major sources of opacity in the visible and near-infrared: H$_2$-H$_2$ collision-induced absorption, Na, K, Rayleigh scattering, TiO and VO (with a condensation curve), H$_2$O, CO, CH$_4$ and CO$_2$ \citep{Christiansen2010, Madhusudhan2010, Madhusudhan2011}. However, given the availability of the {\it Kepler} observation alone in the present work, the number of free parameters far exceed the single data point. Consequently, we explore a range of 1-D $P$-$T$ profiles and chemical composition to explore possible explanations of the observed flux in the \textit{Kepler} bandpass.

Our results indicate that a high geometric albedo is the most plausible explanation of the observed eclipse depth. The major sources of opacity in the {\it Kepler} bandpass are atomic Na and K, molecular TiO and VO where temperatures are higher than the condensation temperature, Rayleigh scattering, and any possible contribution due to clouds and/or hazes. Consequently, we examine the constraints on each of these opacity sources due to the {\it Kepler} point. We assume all other molecules (H$_2$O, CO, CH$_4$ and CO$_2$) to be in chemical equilibrium, assuming solar abundances. We find that the observed {\it Kepler} flux can be explained under three scenarios, shown in the three panels of Fig.~4. 

In a first scenario, we consider a model where all the major absorbers, i.e. Na, K, TiO, and VO, are in chemical equilibrium with solar abundances, and we nominally vary only the $P$-$T$ profile to find a close match to the data. In this scenario, we find that the observed flux cannot be accounted for by thermal emission and Rayleigh scattering alone, both of which together provide a net flux contrast of $\sim 20$ ppm, with Rayleigh scattering contributing a geometric albedo of 0.15. In principle, one might expect that hotter temperature profiles might lead to greater thermal emission which could contribute to the observed flux. However, warmer $P$-$T$ profiles intercept longer absorption columns of TiO/VO which further lower the emergent thermal flux. Therefore, under the assumption of solar abundances of all species, excess flux in the form of reflected light, potentially from clouds and/or hazes, is required to explain the observed {\it Kepler} flux, implying a net geometric albedo of $\sim 0.3$.

In a second scenario, we investigate if thermal emission can be a predominant contributor to the observed {\it Kepler} flux. The strongest absorber in the redder regions of the {\it Kepler} bandpass, where thermal emission dominates, is TiO, followed by VO. As discussed in the equilibrium scenario above, TiO absorption precludes high brightness temperatures due to thermal emission in the {\it Kepler} bandpass. Conversely, we find that thermal emission can contribute substantially to the observed flux, if TiO and VO are assumed to be depleted by over 10$^3$ in the lower atmosphere ($P\sim0.1-1$ bar). Such a scenario just manages to fit the data at the lower error bar, with equal contributions from thermal emission and Rayleigh scattering (with a geometric albedo of 0.17). However, two problems confront this scenario. Firstly, requiring such a high thermal flux implies that all the stellar incident flux on the dayside of the planet must be reradiated on the same side, with almost no energy recirculation to the night side. The {\it Kepler} phase curve of Kepler-7\,b, shown in Fig.~2, could support this scenario if all the flux were purely due to thermal emission. However, because there is still a $\gtrsim$50\% contribution due to Rayleigh scattering required to explain the net flux at occultation, the {\it Kepler} phase curve cannot be interpreted as solely due to a large day-night temperature contrast. Furthermore, at the high pressures ($P \sim 1$ bar) probed by the \textit{Kepler} bandpass, the temperature should be homogenized across day and night side, exhibiting small thermal orbital phase variation. Secondly, at the high temperature of $\sim$2600K probed by \textit{Kepler}, it seems unlikely that TiO/VO can be depleted by factors of $10^3$ below chemical equilibrium at $\sim1$ bar pressure, although the difficulty of sustaining TiO/VO at high altitudes ($P\lesssim10^{-3}$ bar) has been reported in literature \citep{Spiegel2008}.

In our final scenario, we investigate if Rayleigh scattering alone can contribute dominantly to the observed flux. The strongest absorbers of Rayleigh scattered light, as shown in the first two panels of Fig.~4, are atomic Na and K. We find that if we allow depletion of Na and K by a factor of 10-100 of the equilibrium composition, the observed flux in the {\it Kepler} bandpass can be fit extremely well with a geometric albedo of 0.32 from Rayleigh scattering alone along with a nominal contribution due to thermal emission. The resultant model also yields efficient day-night energy circulation.

We finally note that the large planetary radius and very low density would make appealing the hypothesis of Kepler-7\,b being a very young planet that would still be in its cooling phase. However, the asteroseismology results presented in Sect.~2 produce a stellar age of 3.3$\pm$0.4Gyr, which argues for a planetary evolutionary state beyond the collapsing phase \citep{Fortney2005}, and a negligible contribution from internal heat to the occultation depth.

\begin{figure*}
\begin{center}
\epsscale{1.2}
\plotone{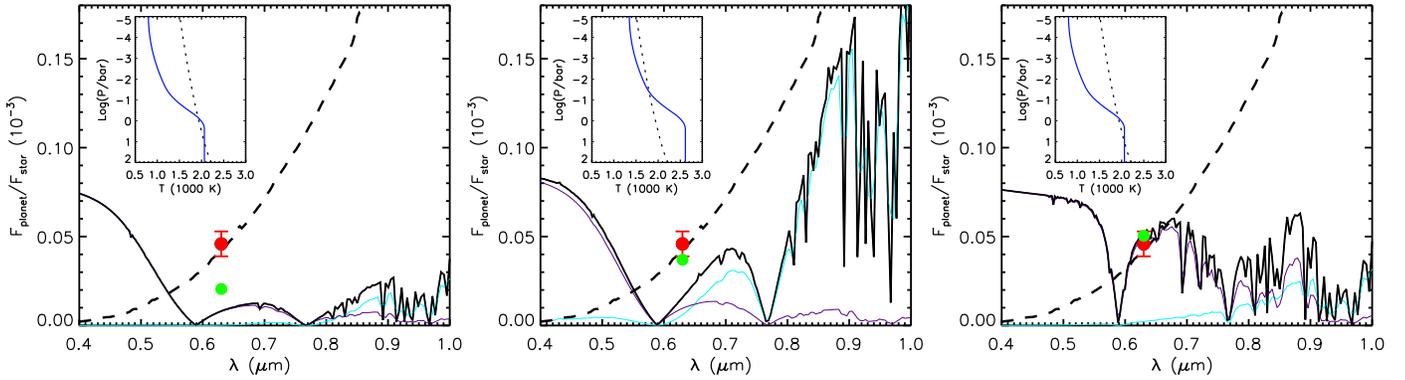}
\caption{Model spectra of Kepler-7\,b in the {\it Kepler} bandpass. Each panel corresponds to each scenario (1 to 3, left to right) described in Sect.4. The {\it Kepler} data point is shown in red. In each panel, the black solid curve shows the net emergent spectrum. The net emergent flux integrated in the {\it Kepler} bandpass is shown in green. The cyan and purple curves show the contributions of thermal emission and reflected light to the emergent spectrum. The black dashed line shows a black body spectrum of the planet at 2550K, divided by a Kurucz model spectrum for the star. The relevant Pressure-Temperature profiles are shown for each atmospheric model in the insets, with the TiO condensation curve in dotted line.\label{fig4}}
\end{center}
\end{figure*}

\section{Conclusions}
Given the three scenarios described in Sect. 4, we interpret the \textit{Kepler} observed planetary flux as due to a combination of Rayleigh scattering and the presence of clouds or a haze layer \citep[e.g.,][]{Lecavelier2008,Sing2009} in the atmosphere of Kepler-7\,b, yielding an averaged geometric albedo of $\sim0.3$ in the {\it Kepler} bandpass. A detailed cloud or haze model is beyond the scope of the present work. Our results motivate new modeling and observational efforts to investigate the nature of clouds and hazes that might be possible in a low gravity atmosphere such as that of Kepler-7\,b.

\acknowledgments
Authors thank the \textit{Kepler} Giant Planet Working Group, P.-O. Quirion, M. Holman, D. Ragozzine, J. Jenkins, J.-M. D\'esert, B. Benneke and D. Latham for useful discussions. We thank E. Dunham, W. Cochran and the referee for helpful comments that improved the manuscript. B.-O.D. acknowledges support from the Swiss NSF and thanks kindly R. Stewart, P.L. Vidale and A. Verhoef from the University of Reading (UK), where part of this work has been carried out. M.G. is Belgian FNRS Research Associate. We acknowledge support from NASA Kepler Participating Science Program NNX10AD67G. Funding for this Discovery Mission is provided by NASA's Science Mission Directorate.

{\it Facilities:}\facility{Kepler},\facility{MMT}

\end{document}